\documentclass[aps,prb,reprint,amsmath,superscriptaddress]{revtex4-1}

\pdfoutput=1

\usepackage{graphicx}
\usepackage{bm}

\begin{document}

\title{Efficient photon detection from color centers in a diamond optical waveguide} 


\author{D. \surname{Le Sage}}
\affiliation{Harvard-Smithsonian Center for Astrophysics, Cambridge, MA 02138, USA}
\affiliation{Physics Department, Harvard University, Cambridge, MA 02138, USA}

\author{L. M. Pham}
\affiliation{School of Engineering and Applied Sciences, Harvard University, Cambridge, MA 02138, USA}

\author{N. Bar-Gill}
\affiliation{Harvard-Smithsonian Center for Astrophysics, Cambridge, MA 02138, USA}
\affiliation{Physics Department, Harvard University, Cambridge, MA 02138, USA}

\author{C. Belthangady}
\affiliation{Harvard-Smithsonian Center for Astrophysics, Cambridge, MA 02138, USA}
\affiliation{Physics Department, Harvard University, Cambridge, MA 02138, USA}

\author{M. D. Lukin}
\affiliation{Physics Department, Harvard University, Cambridge, MA 02138, USA}

\author{A. Yacoby}
\affiliation{Physics Department, Harvard University, Cambridge, MA 02138, USA}

\author{R. L. Walsworth}
\email{rwalsworth@cfa.harvard.edu}
\affiliation{Harvard-Smithsonian Center for Astrophysics, Cambridge, MA 02138, USA}
\affiliation{Physics Department, Harvard University, Cambridge, MA 02138, USA}

\date{\today}

\begin{abstract}
A common limitation of experiments using color centers in diamond is the poor photon collection efficiency of microscope objectives due to refraction at the diamond interface. We present a simple and effective technique to detect a large fraction of photons emitted by color centers within a planar diamond sample by detecting light that is guided to the edges of the diamond via total internal reflection. We describe a prototype device using this ``side-collection'' technique, which provides photon collection efficiency $\approx47\%$ and photon detection efficiency $\approx39\%$. We apply the enhanced signal-to-noise ratio gained from side-collection to AC magnetometry using ensembles of nitrogen-vacancy (NV) color centers, and demonstrate AC magnetic field sensitivity $\approx100\:\rm{pT/\sqrt{Hz}}$, limited by added noise in the prototype side-collection device.  Technical optimization should allow significant further improvements in photon collection and detection efficiency as well as sub-picotesla NV-diamond magnetic field sensitivity using the side-collection technique.
\end{abstract}


\maketitle

\begin{figure}[!]
\includegraphics{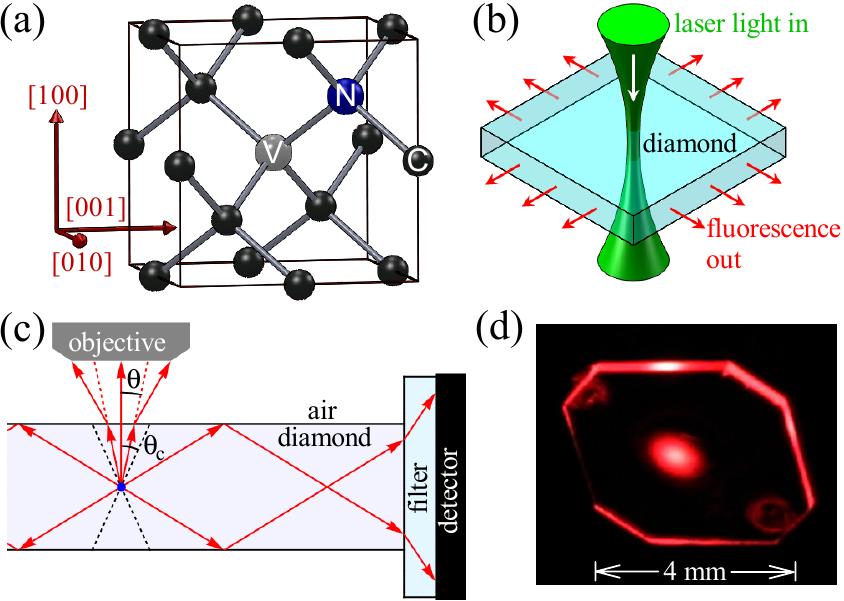}
\caption{\label{fig:fig1}(Color online) (a) The NV color center consists of a substitutional nitrogen atom (N) adjacent to a vacancy (V) in the diamond lattice. (b) In the side-collection technique, a focused laser beam excites color centers in a specific volume within the diamond, and much of the resulting fluorescence is detected after it exits one or more sides of the diamond waveguide. In the demonstration experiments reported here, four photodetectors are arranged around the four primary sides of the diamond chip. (c) Ray diagrams illustrate refraction at the diamond surface and light guided by total internal reflection above the critical angle ($\theta_{c}=24.6^{\circ}$). (d) Red-filtered photograph of NV fluorescence from the diamond chip used in the demonstrations reported here. Guiding of NV fluorescence light is evident as a bright glow around the diamond's perimeter, while a 532 nm laser beam passes through its center.}
\end{figure}

A wide variety of fluorescent point defect centers in diamond have been identified, with possible applications toward quantum information and biological imaging.\cite{Aharonovich2011} One such color center that has attracted considerable attention is the negatively-charged nitrogen-vacancy center (NV) [Fig.~\ref{fig:fig1}(a)], which has an electronic structure that allows for optical initialization and detection of the electronic spin, and coherent manipulation of the spin state, \cite{Jelezko2004} with long room-temperature spin coherence times.\cite{BalasubramanianNatMat2009} Demonstrated applications of NVs include single photon generation,\cite{Beveratos2001,Babinec2010} quantum information processing, \cite{Dutt2007,Jiang2009,Togan2010} super-resolution microscopy,\cite{Rittweger2009,Maurer2010} nanoscale magnetometry using a single NV, \cite{MazeNature2008,Balasubramanian2008} and vector magnetic field imaging using ensembles of NVs.\cite{Pham2011} For all of these applications, the photon collection efficiency is critically important.

Conventionally, photons emitted from NVs in a bulk diamond substrate are collected using a microscope objective with a large numerical aperture (NA). However, refraction at the flat diamond interface reduces the effective NA of the objective by a factor equal to the refractive index of diamond ($\rm{n_{d}=2.4}$), resulting in a photon collection efficiency ($\eta_{c}$) $\leq 8\%$.\cite{SupMat} Photon detection efficiencies ($\eta_{d}$) $<2\%$ are typical when detector coupling and quantum efficiencies are taken into account.

We note that NVs in diamond nanostructures can avoid the $\eta_{c}$-degradation caused by a bulk diamond interface. For example, NVs in subwavelength-size diamond nanocrystals can be approximated as point emitters in the medium surrounding the nanocrystal, yielding higher $\eta_{c}$ but lower emission rates due to increased excited state lifetime.\cite{Beveratos2001} Light from NVs inside fabricated diamond nanowires is primarily emitted along the nanowire symmetry axis, allowing $\eta_{c}\approx40\%$.\cite{Babinec2010} However, NVs in diamond nanostructures typically have much shorter spin coherence times \cite{Tisler2009} than those in bulk diamond.\cite{BalasubramanianNatMat2009,Stanwix2010} One approach to increase $\eta_{c}$ for NVs in bulk diamond is to construct a solid-immersion lens (SIL) out of the diamond substrate, using either macroscopic \cite{Siyushev2010} or microscopic \cite{Hadden2010} fabrication techniques. Light from NVs at the center of a hemispherical diamond SIL passes through the surface at normal incidence, allowing $20\%<\eta_{c}<30\%$. However, this improved photon collection efficiency is limited to NVs lying $\rm{<30\:\mu m}$ from the center of a macroscopic SIL.\cite{Siyushev2010} For a larger field of view, or for large-volume ensemble measurements in bulk diamond, a different collection-enhancement technique is required.

Here we demonstrate an efficient fluorescence detection technique that uses the planar diamond chip containing the color centers as an optical waveguide. For this ``side-collection'' technique, photons emitted by the color centers are confined between the two parallel diamond chip surfaces by total internal reflection (TIR) and guided to the sides (edges) of the diamond chip, where they pass through a filter onto a detector [Fig.~\ref{fig:fig1}(b,c)]. This technique avoids complicated fabrication procedures and allows efficient photon detection from color centers located anywhere within the diamond volume.

We estimated the expected efficiency of the side-collection technique using theoretical models of the average NV emission pattern and realistic approximations of the diamond chip geometry and acceptance angles of the detectors.\cite{SupMat} We find that $\rm{\approx91\%}$ of the NV fluorescence is confined by TIR between the polished (100) planar surfaces of the diamond chip, and $\rm{\approx29\%}$ reaches the detectors on first incidence of the photons with the diamond chip sides (edges). Much of the light also undergoes TIR off the sides, but may reach the detectors after many reflections within the diamond. We therefore expect $29\%<\eta_{c}<91\%$ for the side-collection technique, depending on details of the experimental geometry, diamond chip surface properties, etc.

\begin{figure}
\includegraphics{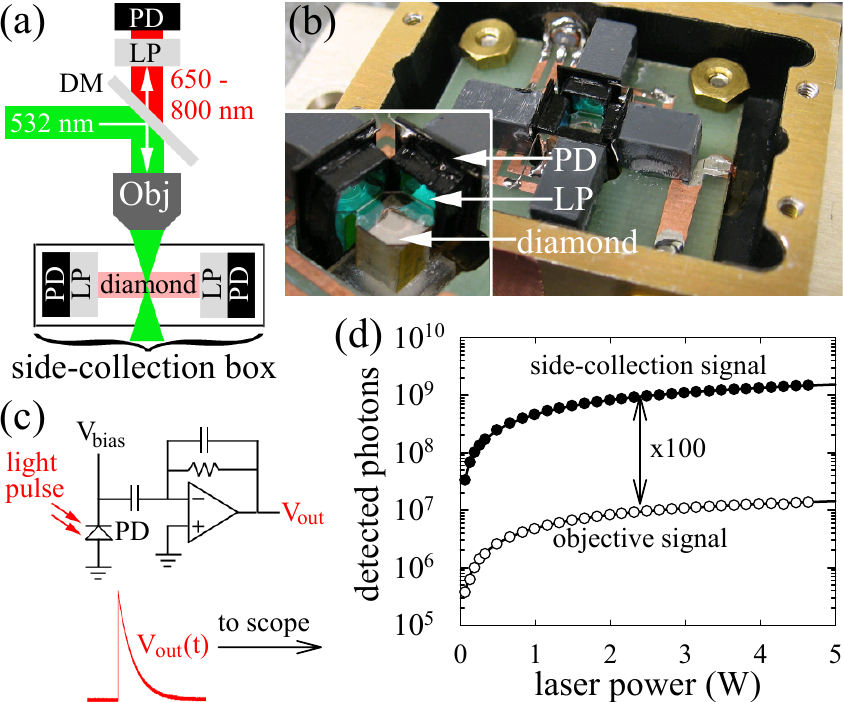}
\caption{\label{fig:fig2}(Color online) (a) Prototype device to compare the photon collection and detection efficiency of the side-collection technique with that of a 0.4-NA microscope objective (Obj), using a high NV-density diamond chip, 532 nm excitation laser, dichroic mirror (DM), 650 nm long-pass filter (LP), and Si photodiodes (PD). (b) Photo of the side-collection prototype device (partially disassembled for viewing of the interior). The inset shows the locations of the LPs and PDs in relation to the diamond chip. (c) Side-collection and microscope objective optical signals were recorded using a charge-sensitive amplifier. (d) NV fluorescence (number of photons) detected by the two collection modalities during a 300 ns pulse of the excitation laser (532 nm), as a function of laser power, shows a 100-times larger side-collection signal.}
\end{figure}

We constructed a prototype device to compare the side-collection technique to the conventional microscope objective method [Figs.~\ref{fig:fig2}(a)-\ref{fig:fig2}(c)]. This prototype instrument employed a [100]-oriented diamond chip ($\rm{4.3\: mm}$ $\times$ $\rm{4.3\: mm}$ $\times$ $\rm{0.2\: mm}$) grown via chemical vapor deposition (CVD) with a high NV-density $\rm{\sim10^{15}\:cm^{-3}}$ (Apollo Diamond, Inc.) [Fig.~\ref{fig:fig1}(d)]. Four rectangular 650 nm longpass optical filters were placed in contact with the edges of the diamond. The filters transmitted most of the NV fluorescence band ($\rm{\sim637-800\:nm}$), while reflecting scattered 532 nm excitation light. Four chip-style Si photodiodes were affixed to the backs of the filters. Because the filters were on a 2 mm thick quartz substrate, large active-area photodiodes (6 mm x 7 mm) were employed to maximize the detection acceptance angle. Future designs may be simplified using reflective coatings on several diamond surfaces and integrated filters on the photodiodes.

We calibrated the prototype device's photon side-collection efficiency by comparing it against a microscope objective with known $\eta_{c}$. Light from a 532 nm laser was focused into the diamond sample by a microscope air-objective (NA=0.40) [Fig.~\ref{fig:fig2}(a)]. The objective also collected NV fluorescence, and directed it to a filter and photodiode identical to those used for side-collection. The objective's low NA guaranteed that its line-of-sight was not obstructed by the side-collection filters. In Fig.~\ref{fig:fig2}(d) we compare the integrated number of photons detected with the two collection modalities during a $\rm{300\:ns}$ laser pulse, at various laser powers. The integrated photon count was measured by alternately connecting a charge-sensitive amplifier (Cremat Inc. CR-112) to the photodiode(s) of the two collection paths, and recording the average signal amplitude [Fig.~\ref{fig:fig2}(c)]. We found that the side-collection signal had a $100\pm5$ times larger photon count than the microscope objective signal under identical experimental conditions. The theoretically-estimated collection efficiency for the microscope objective was 0.59\%.\cite{SupMat} Transmission losses through the objective, dichroic, and filter reduced the fraction of light reaching the detector to $\eta_{c}\approx0.47\%$. The average quantum efficiency of the photodiode within the NV emission band at near-normal incidence $\approx 83\%$, indicating $\eta_{d}\approx0.39\%$. This implies that the NV fluorescence detection efficiency of the side-collection prototype device $\approx39\%$, while the fraction of photons reaching the four side-collection photodiodes (over a wide range of incidence angles) was $\geq47\%$. We also confirmed that the optical signals in the two detection paths were due to NV fluorescence by sweeping the frequency of a microwave field across the characteristic NV electron spin resonance at 2.87 GHz.\cite{Gruber1997} Both detection paths had a typical and nearly identical signal response, indicating that the side-collection filters selected NV fluorescence as effectively as conventional fluorescence microscopy.

\begin{figure}
\includegraphics{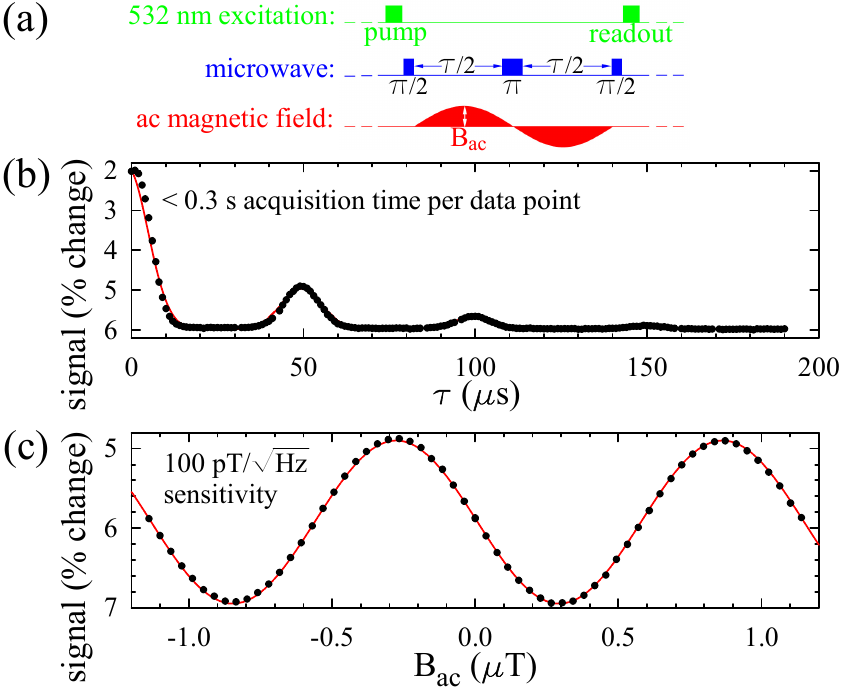}
\caption{\label{fig:fig3}(Color online) (a) Optical and microwave spin-echo pulse sequence used for NV-diamond AC magnetometry with the side-collection prototype device.\cite{MazeNature2008} An optical pump pulse prepares the NV spins in the $\rm{|m_{s} = 0>}$ state; a microwave spin-echo sequence allows the NV spins to probe an applied 20 kHz magnetic field; and an optical readout pulse allows measurement of the change in NV spin-dependent fluorescence compared to the initial state. (b) By varying $\tau$ with $\rm{B_{ac}=0}$, we measure the NV spin-echo decoherence curve shown, with greatly enhanced signal-to-noise ratio provided by the side-collection technique. (c) Varying $\rm{B_{ac}}$ with $\tau = 50\;\mu\rm{s}$ produces the NV-diamond AC magnetometry curve shown, yielding $\rm{100\:pT/\sqrt{Hz}}$ sensitivity for a laser excitation volume $\rm{\sim10^{-4}\: mm^{3}}$.}
\end{figure}

One important application of the side-collection technique is sensitive magnetometry using ensembles of NV spins in a diamond chip.\cite{TaylorNatPhys2008,Pham2011} To demonstrate this application using the side-collection prototype device, we performed AC magnetometry with $\sim10^{7}$ NVs contributing to the magnetometry signal in a $\rm{\sim10^{-4}\: mm^{3}}$ laser excitation volume, using the standard spin-echo technique described in Ref.~\onlinecite{MazeNature2008} [Fig.~\ref{fig:fig3}(a)]. We applied a 37.5 G static magnetic field in the [111] direction, so that we could use an applied microwave field to resonantly select $\rm{|m_{s} = 0>\leftrightarrow |m_{s}=1>}$ spin transitions of [111]-oriented NVs, as in Ref.~\onlinecite{Pham2011}. (1/4 of the total number of NV spins are resonantly manipulated with this method, as NV orientations are distributed equally among the four crystallographic axes.) By sweeping the spin-echo duration ($\tau$), we generated a  $\rm{T_{2}}$ coherence curve with revivals at even multiples of the $\rm{^{13}C}$ nuclear precession period \cite{MazeNature2008} [Fig.~\ref{fig:fig3}(a)]. We extracted $\rm{T_{2}=35\:\mu s}$ for this high NV-density diamond sample. We then generated the magnetometry curve in [Fig.~\ref{fig:fig3}(b)] by varying the amplitude of a 20 kHz AC magnetic field, with $\rm{\tau=50\:\mu s}$. From the slope of the curve and the standard deviation in the measurements at zero AC field amplitude, we found a magnetic field sensitivity $\rm{\approx100\:pT/\sqrt{Hz}}$, which to our knowledge is the best magnetic field sensitivity demonstrated to date with any NV-diamond system.\cite{BalasubramanianNatMat2009,Acosta2010} Note that this magnetic field sensitivity was limited by electrical noise in the detection circuitry of the prototype side-collection device, as verified by independent measurements. We expect that this added noise can be significantly reduced in future, optimized side-collection instruments, enabling greatly improved magnetic field sensitivity. For example, if the side-collection measurements for the current NV-diamond chip were instead limited by photon shot noise ($2\times10^{8}$ photons detected per measurement), then the magnetic field sensitivity would be $\rm{\approx4\:pT/\sqrt{Hz}}$. Sensitivities below $\rm{1\:pT/\sqrt{Hz}}$ could be achieved by using low-noise detection circuitry, increasing the measurement volume, and lengthening the NV $\rm{T_2}$ through diamond engineering \cite{BalasubramanianNatMat2009} or dynamic decoupling techniques.\cite{Naydenov2011}

The photon detection enhancement of the side-collection technique has many other potential applications. For example, magnetometry may be extended to magnetic field imaging by using a thin layer of NVs near the diamond chip surface \cite{Pham2011} and scanning the laser focus while making time-resolved side-collection fluorescence measurements. The high signal-to-noise ratio provided by side-collection also greatly increases the speed of NV ensemble measurements, which may be used to study decoherence processes and develop NV spin manipulation protocols for magnetometry and quantum information. For example, the NV decoherence and magnetometry data shown in Fig.~\ref{fig:fig3} was generated with $<0.3\:\rm{s}$ of signal averaging time per data point, which is orders of magnitude shorter than the time that would be required to achieve comparable signal-to-noise using a microscope objective. The side-collection technique should also be applicable to single NV measurements using a tightly-focused laser beam for NV excitation. However, because side-collection accepts out-of-focus light, an ultra-pure diamond would likely be required to isolate single NVs in the excitation volume. This condition becomes less restrictive if a stimulated emission depletion (STED) microscope is employed,\cite{Rittweger2009} and we note that side-collection could increase the speed and precision of such super-resolution techniques, including combined NV super-resolution imaging and magnetometry (mag-STED).\cite{Maurer2010} Furthermore, the higher photon count rate provided by side-collection would improve NV quantum-state readout fidelity, and may be used in conjunction with existing techniques to achieve single-shot readout of the NV spin for quantum information applications.\cite{Jiang2009}
\\

We gratefully acknowledge the provision of diamond samples by Apollo Diamond and Element Six; and also informative discussions with Patrick Doering, David Glenn, and Alexei Trifonov. This work was supported by NIST, NSF, and DARPA (QuEST and QuASAR programs).

\bibliography{side-collection_arxiv}

\end{document}